\documentclass{iau-JDSS}

\def\simgt{\lower.5ex\hbox{$\; \buildrel > \over \sim \;$}}
\def\simlt{\lower.5ex\hbox{$\; \buildrel < \over \sim \;$}}
\def\Porb{$P_{\rm orb}$}
\def\Pbif{$P_{\rm bif}$}
\def\Chandra{${\it Chandra}$}
\def\HST{${\it HST}$\ }
\def\arcsec{\hbox{$^{\prime\prime}$}}

\newcommand{\kms}{$\rm {km}\, \rm s^{-1}$}

\title[Neutron stars and black holes in star clusters] 
{Neutron stars and black holes in star clusters}

\author[Baumgardt \etal\ ]   
{F.~A.~Rasio$^1$, H.~Baumgardt$^2$, A.~Corongiu$^3$, F.~D'Antona$^4$, G.~Fabbiano$^5$, J.~M.~Fregeau$^1$, K.~Gebhardt$^6$, C.~O.~Heinke$^1$, P.~Hut$^7$, N.~Ivanova$^8$, T.~J.~Maccarone$^9$, S.~M.~Ransom$^{10}$, 
N.~A.~Webb$^{11}$}

\affiliation{$^1$Northwestern University, Dept of Physics and Astronomy, Evanston, Illinois, USA \\[\affilskip]
$^2$Argelander Institut f\"ur Astronomie, University of Bonn, Germany \\[\affilskip]
$^3$INAF, Osservatorio di Cagliari e Universit\`a di Cagliari, Italy \\[\affilskip]
$^4$Osservatorio Astronomico di Roma, Monteporzio, Italy \\[\affilskip]
$^5$Harvard-Smithsonian Center for Astrophysics, Cambridge, Massachusetts, USA \\[\affilskip]
$^6$Astronomy Department, University of Texas at Austin, USA \\[\affilskip]
$^7$Institute for Advanced Study, Princeton, New Jersey, USA
\\[\affilskip]
$^8$Canadian Institute for Theoretical Astrophysics, Toronto, Ontario, Canada
\\[\affilskip]
$^{9}$School of Physics and Astronomy, University of 
                        Southampton, UK \\[\affilskip]
$^{10}$NRAO, Charlottesville, Virginia, USA \\[\affilskip]
$^{11}$Centre d'Etude Spatiale des Rayonnements, Toulouse, France}

\pubyear{2006}
\volume{Volume 14}  
\pagerange{page--page}
\date{11/1/06}
\jname{Highlights of Astronomy, Volume 14}
\editors{K.A.\ van der Hucht, ed.}

\begin{document}

\maketitle

\firstsection 
              
\section{Introduction}
This article was co-authored by all invited speakers at the Joint Discussion
on ``Neutron Stars and Black Holes in Star Clusters,'' which took place
during the IAU General Assembly in Prague, Czech Republic, on August~17
and~18, 2006. Each section presents a short summary of recent developments
in a key area of research, incorporating the main ideas expressed during the
corresponding panel discussion at the meeting.

Our meeting, which had close to 300 registered participants,
was broadly aimed at the large community of
astronomers around the world working on the formation and evolution
of compact objects and interacting binary systems in dense star clusters, 
such as globular clusters and galactic nuclei. The main scientific topics 
cut across all traditional boundaries, including Galactic and
extragalactic astronomy, environments from young starbursts to old globular
clusters, phenomena from radio pulsars to gamma-ray bursts, and
observations using ground-based and space-based telescopes, with a
significant component of gravitational-wave astronomy and relativistic
astrophysics.

Great advances have occurred in this field during the past few
years, including the introduction of fundamentally new theoretical paradigms
for the formation and evolution of compact objects in binaries as well as
countless new discoveries by astronomers that have challenged many accepted
models. Some of the highlights include: a nearly complete census of all the
millisecond pulsars in 47~Tucanae; first detections of many new radio pulsars in
other clusters, particularly Terzan~5;
detailed studies of X-ray binary populations and their luminosity
functions in many galaxies and extragalactic globular clusters; increasing evidence for
intermediate-mass black holes in clusters and greatly improved theoretical
understanding of their possible formation processes.

The next few years will prove at least as exciting, with many more
data sets coming from recently or soon-to-be
launched satellites, many new objects found 
in extensive deep radio and X-ray surveys, and follow-up
spectroscopy and photometry with optical telescopes. On the theoretical side,
advances in computer codes and special-purpose hardware will allow for more
and more realistic modeling of whole large clusters including fairly complete
treatments of all the relevant physics.

\section{Direct $N$-body Simulations}

Direct $N$-body simulations follow stars individually.  This is
important when modeling star clusters, where specific interactions
between single stars and binaries, as well as more complex multiple
systems, play a central role (\cite[Aarseth 2003]{aar03};
\cite[Heggie \& Hut 2003]{heghut03}).  In contrast, for larger-scale
simulations of encounters between galaxies, and cosmological
simulations in general, the stars are modeled as a fluid in phase
space, and the individual properties of the stars are no longer
important.

Traditionally, the term ``collisionless stellar dynamics'' has been used
for the latter case, and ``collisional stellar dynamics'' for the former
case.  When these terms were coined in the nineteen sixties, they were
perhaps appropriate, but now that we have started to simulate physical
collisions between stars in a serious way, the use of the word
``collision'' in the old sense has become rather confusing, since it was
meant to denote relatively distant encounters that contribute to the
two-body relaxation of a system.

It may be useful to introduce a new expression for the study of
star clusters, shorter than ``collisional stellar dynamics,'' and
broader in the sense of including stellar evolution and hydrodynamics
as well.  One option would be to use {\it smenology\/}, or in Greek
$\sigma \mu \eta \nu o \lambda o \gamma \iota \alpha$, after
$\sigma \mu \eta \nu o \sigma$ ({\it smenos\/}), swarm, which is the
word in use in modern Greek for a cluster; a star cluster is called
$\sigma \mu \eta \nu o \sigma$ $\alpha \sigma \tau \epsilon \rho \omega \nu$
({\it smenos asteroon\/}), literally a swarm of stars\footnote{Dimitrios
Psaltis, personal communication.}.

Simulations of dense stellar systems, such as globular clusters (hereafter GCs) and
galactic nuclei, have never yet been very realistic.  Simplifying
assumptions, such as those used in gas models or Fokker-Planck and Monte
Carlo codes, have allowed us to model large particle numbers at the
expense of a loss of detail in local many-body interactions and the
imposition of global symmetry constraints.  Conversely, direct $N$-body
integration, while far more accurate, has labored under a lack of
computer speed needed to model a million stars.

The good news is that we will soon be approaching effective computer
speeds in the Petaflops range (\cite[Makino 2006]{mak06}), which will
allow us to model the gravitational million-body problem with full
realism, at least on the level of point particles.  Adding equally
realistic stellar evolution and hydrodynamics will be no problem as
far as the necessary computer speed is concerned.

When the hardware bottleneck will thus be removed, the software
bottleneck for realistic cluster simulations will become painfully
obvious (\cite[Hut 2007]{hut07}).  This is the bad news.  While some
serious uncertainties remain in the science needed to improve the
software, currently the main bottleneck is neither science nor
computer speed, but rather a sufficiently robust implementation of
already available knowledge.

The main two codes currently being used for direct $N$-body simulations, {\tt NBODY4} and
{\tt Kira}, are both publicly available:
{\tt NBODY4} at {\tt www.ast.cam.ac.uk/\~{}sverre/web/ pages/nbody.htm} and
{\tt Kira} at {\tt www.ids.ias.edu/\~{}starlab/}.

{\tt NBODY4} and other related codes form the results of a
more than forty-year effort by Sverre Aarseth, as documented in \cite{aar03}.
These codes are written in Fortran and they can be run stand-alone.
A parallel version has been developed, named
{\tt NBODY6++} (\cite[Spurzem 1999]{spu99}; \cite[Spurzem \& Baumgardt 2003]{spubau03}), publicly available at
{\tt ftp://ftp.ari. uni-heidelberg.de/pub/staff/spurzem/nb6mpi/}.
In addition to stellar dynamics, the version of {\tt NBODY4}
developed by Jarrod Hurley and collaborators (see \cite[Hurley \etal\ 2005]{hur05}
and references therein) includes a treatment of stellar evolution for
both single stars (named {\tt SSE}) and binary stars (named {\tt BSE}), using
fitting formulae and recipes (\cite[Hurley \etal\ 2002]{hur02}).

The {\tt Kira} code forms an integral part of the {\tt Starlab} environment
(\cite[Portegies Zwart \etal\ 2001]{por01}).  {\tt Kira} and {\tt Starlab}
 are written in C++.  The basic data
structure of {\tt Kira} consists of a flat tree containing leaves
representing single stars as well as nodes that hold center of mass
information for small clumps of interacting stars.  Each clump is
represented by a binary tree, where each node determines a local
coordinate system.  The {\tt Kira} code has built-in links to {\tt Seba}, a
stellar evolution module using fitting formulae developed by
\cite{tou96} and recipes developed by \cite{porver96}.  In addition to
{\tt Kira} and {\tt Seba}, the {\tt Starlab} environment contains tools for setting up
initial conditions for star clusters, using various models, and for
analyzing the results of $N$-body simulations.  {\tt Starlab} also contains
packages for binary--single-star and for binary--binary scattering.

Within the next ten years, multi-Petaflops computers will enable us
to follow the evolution of star clusters with up to a million stars
(\cite[Makino 2006]{mak06}).  To make efficient use of this opportunity,
while including increasingly realistic treatments of stellar evolution
and stellar hydrodynamics, a number of new developments are required.

On the purely stellar dynamics level, some form of tree code may be
useful for speeding up the long-range force calculations, as pioneered
by \cite{mcmaar93}.  In addition, guaranteeing accurate treatments of
local interactions will become more challenging, especially for
extreme mass ratios; designing good algorithms for following the
motions of stars in the neighborhood of a massive black hole is
currently an area of active research.

The largest challenge, however, will be to develop robust stellar
evolution and stellar hydrodynamics codes, that can interface reliably
with stellar dynamics codes, without crashing.  The MODEST initiative
(for MOdeling DEnse STellar systems) was started in 2002 with the
intention to provide a forum for discussions concerning this challenge
(\cite[Hut \etal\ 2003]{hut03}). A pilot project, MUSE (for
MUlti-scale MUlti-physics Scientific Environment), was initiated recently
to develop a
modular software environment for modeling dense stellar systems,
allowing packages written in different languages to interoperate
within an integrated software framework (see the MODEST web site at
{\tt www.manybody.org/modest.html} and click on ``projects'').

Finally, for any large software project that involves a team of code
developers, good documentation is essential.  For most astrophysical
simulation codes, documentation has come mainly as an afterthought.
An attempt to develop a new code for modeling dense stellar systems,
using an almost excessive amount of documentation can be found at
{\tt www.ArtCompSci.org}, the web site for ACS (the Art of
Computational Science).

\section{Monte Carlo Methods}\label{sec:fregeau}

H\'enon's Monte Carlo method has given rise to an industry in the business of simulating the
evolution of dense stellar systems, providing fast and accurate simulations of large-$N$ systems.
Its computational speed, coupled with the physical assumptions it requires (notably spherical symmetry
and dynamical equilibrium) make it a very natural complement to ``direct'' $N$-body simulations
(Sec.~2), 
which are computationally much more expensive (or, equivalently, allow for smaller $N$) and generally
require the use of special-purpose (GRAPE) hardware.  Here we briefly discuss
the method and the primary Monte Carlo codes before focusing on recent progress on the technique
and recent contributions to the study of dense stellar systems made by those codes.

At the core of the modern Monte Carlo method is the H\'enon technique, which amounts to the following.
The evolution of each particle's orbit in a dense stellar system is influenced by all
other particles in the system, although one cannot realistically sum all two-body scattering interactions
and achieve reasonable computational speed on standard hardware.  Instead, for each particle one performs a 
``super''-encounter with a nearby particle, with the deflection angle chosen so as to represent the effects
of relaxation due to the whole cluster (see Freitag \& Benz 2001 for a pedagogical discussion and original
references).  This ``trick'' makes the Monte Carlo method scale with particle number as $N \log N$ per
time step, instead of $N^2$ with direct $N$-body methods.

There are currently three primary Monte Carlo evolution codes in use, along with a fourth hybrid 
gas model/Monte Carlo code.  The Northwestern code uses the H\'enon technique with a timestep
shared among all particles (Fregeau \& Rasio 2006, and references therein).  
Notably, it incorporates dynamical integration of all binary
interactions in a cluster, allowing for the study of the long-lived binary-burning phase.
The Freitag code uses the H\'enon technique with a radius-dependent timestep 
(Freitag \etal\ 2006b, and references therein).  It is notable
for its inclusion of physical stellar collisions drawn from a library of SPH simulations, as well
as a treatment of loss cone physics.  The Giersz code uses the H\'enon technique
with radial timestep zones (Giersz 2006, and references therein), and includes stellar
evolution of single stars.  The Giersz \& Spurzem hybrid code couples a gas dynamical model
for the single star population with a Monte Carlo treatment of binary interactions 
(Giersz \& Spurzem 2003, and references therein).  Table~\ref{tab:mc} lists the capabilities
of the Monte Carlo codes just discussed, as well as those of the direct $N$-body method.

\begin{table}\def~{\hphantom{0}}
  \begin{center}
    \caption{Comparison of the capabilities of different methods for simulating the evolution of dense
      stellar systems.  The first column lists the different physical processes at
      work in  stellar systems, column ``NB'' lists the capabilities of the $N$-body method, 
      column ``MC'' lists what the Monte Carlo method is in principle capable of, columns ``NU,''
      ``F,'' ``G,'' and ``GS,'' list the current capabilities of the Northwestern, Freitag, and Giersz
      Monte Carlo codes, as well as the Giersz \& Spurzem hybrid gas/Monte Carlo code.  A filled circle
      means the code is fully capable of treating the physical process, while an open circle means it 
      is capable subject to some limitations.}
    \label{tab:mc}
    \begin{tabular}{ccccccc}
      \hline
          {\bf Physics} & {\bf NB} & {\bf MC} & {\bf NU} & {\bf F} & {\bf G} & {\bf GS}\\
          \hline
          two-body relaxation & $\bullet$ & $\bullet$ & $\bullet$ & $\bullet$ & $\bullet$ & $\bullet$ \\
          stellar evolution & $\bullet$ & $\bullet$ & $\circ$ & $\circ$ & $\circ$ & \\
          stellar collisions & $\bullet$ & $\bullet$ & $\bullet$ & $\bullet$ & & \\
          binary interactions & $\bullet$ & $\bullet$ & $\bullet$ & & $\circ$ & $\bullet$ \\
          external effects & $\bullet$ & $\circ$ & $\circ$ & $\circ$ & $\circ$ & $\circ$ \\
          central BH & $\bullet$ & $\bullet$ & & $\bullet$ & & \\
          rotation & $\bullet$ & & & & & \\
          violent relaxation & $\bullet$ & & & & & \\
          large-angle scattering & $\bullet$ & $\bullet$ & & & & \\
          three-body binaries & $\bullet$ & $\bullet$ & & & $\bullet$ & $\bullet$ \\
          large $N$, $f_b$ & & $\bullet$ & $\bullet$ & $\bullet$ & $\bullet$ & $\bullet$ \\
          \hline
    \end{tabular}
  \end{center}
\end{table}

In the past year or so, six papers relying on the Monte Carlo codes mentioned
above have been published.  G\"urkan \etal\ (2006) studied the process of 
runaway collisional growth in young dense star clusters, and found the exciting new result
that, when the runaway process operates in young clusters with primordial binaries, 
generally two very massive stars (VMSs) are formed.  The VMSs formed may quickly
undergo collapse after formation to become intermediate-mass black holes (IMBHs), yielding the exotic
possibility of IMBH--IMBH binaries forming in young clusters.  Giersz (2006) performed
simulations of clusters with $N=10^6$ stars subject to the tidal field of their parent galaxy.
The large particle number allowed a detailed study of the evolution of the cluster
mass function.  Freitag \etal\ (2006c) performed a comprehensive study of
the process of mass segregation in galactic nuclei containing supermassive black holes, with
implications for the distribution of X-ray binaries (XRBs) at the Galactic center.  
Freitag \etal\ (2006a,b) studied in great detail the process of runaway collisional growth in young dense star
clusters.  Their study yielded several key results.  First, a comparison
of approximate physical stellar collision prescriptions with the detailed results
of SPH simulations showed that the simple ``sticky-star'' approximation---in which
stars are assumed to merge without mass loss when their radii touch---is sufficiently accurate 
for clusters with
velocity dispersions less than the typical stellar surface escape velocity to faithfully model
the physics of runaway collisions.  Second, runaway collisional growth of a VMS to
$\sim 10^3 \,M_{\odot}$ is generic for clusters with central relaxation times sufficiently
short ($\lesssim 25\,{\rm Myr}$) and for clusters which are initially collisional.  
Fregeau \& Rasio (2006) presented the first Monte Carlo simulations of clusters
with primordial binaries to incorporate full dynamical integration of binary scattering 
interactions (the work of Giersz \& Spurzem 2003 performed integration of binary interactions, 
but used a gas dynamical model for the single star population).  They performed detailed 
comparisons with direct $N$-body calculations, as well as with semi-analytical theory for cluster
core properties as a function of the binary population, and found good agreement with both.
They then simulated an ensemble of systems and compared the resulting cluster structural parameters
($r_c/r_h$ and the concentration parameter, for example) during the binary burning phase with the observed
Galactic GC population.  The interesting result is that the values of $r_c/r_h$
predicted by the simulations are roughly a factor of 10 smaller than what is observed.  
The most likely explanation is that physical processes ignored in the simulations (such as stellar evolution and
collisions) are at work in the Galactic sample, generating energy in the cores and causing them
to expand.  However, more detailed simulations should be performed to test this hypothesis.

\section{Stellar and Binary Evolution in Globular Clusters}

Much of the recent work in the area of stellar evolution in GCs
 has concentrated on the evolution of low-mass
X-ray binaries (LMXBs) and their likely remnants, the millisecond pulsars
(MSPs). In particular, several new studies have considered the possible effects
of a ``radio--ejection phase'' initiated when the mass transfer temporarily stops 
during the secular evolution of the systems. 

The much larger fraction of binary MSPs and LMXBs in
GCs, with respect to their fraction in the Galactic field, is 
regarded as a clear indication that binaries containing neutron stars (NSs) in GCs are
generally not primordial, but are a result of stellar encounters due to
the high stellar densities in the GC cores. On the other hand, it is still not
clear how the LMXBs are formed in the Galactic field, as the result of a supernova
explosion in a binary in which the companion is a low-mass star will generally 
destroy the binary. Many possible processes have been invoked to explain LMXBs: (i) 
accretion-induced collapse of a white dwarf primary into a neutron star; 
(ii) supernova kicks due to asymmetric neutrino energy deposition 
during the supernova event; (iii) formation of LMXBs as remnants of 
the evolution of binaries with intermediate-mass donors; (iv) LMXBs formed 
by capture in the dense environment of GCs and later released when the GC is
destroyed.
This last hypothesis, originally due to Grindlay (1984), was recently re-evaluated in
the literature (Podsiadlowski \etal\ 2002; see also Sec.~11). 

The ``standard'' secular evolution of LMXBs as progenitors of binary MSPs is 
reasonable in the context of the evolution of binaries above the so-called 
``bifurcation period'' \Pbif\ (Tutukov \etal\ 1985; Pylyser \& Savonije 1988), in which
the donor star begins the mass transfer phase after it has finished the
phase of core hydrogen burning, and the system ends up as a low-mass white
dwarf (the remnant helium core of the donor) in a relatively long
or very long period orbit with a radio MSP (e.g., Rappaport \etal\ 1995).
Some of these systems may also be the remnants of the evolution 
of intermediate-mass donors, with similar resulting orbital periods (Rasio \etal\ 2000; 
Podsiadlowski \etal\ 2002). Recently, \cite{dantona2006} showed that the secular evolution 
at P$>$\Pbif\ may need to take into account the detailed stellar evolution
of the giant donor, in order to explain the orbital period gap of binary MSPs between
$\sim 20$\ and $\sim 60$ days. During the evolution along the RGB, the 
hydrogen burning shell encounters the hydrogen chemical discontinuity 
left by the maximum deepening of convection: 
the thermal readjustment of the shell causes a luminosity and radius drop, which
produces a well known ``bump'' in the luminosity function of the RGB
in GCs. In semi-detached binaries, at the bump, the mass transfer is 
temporarily stopped, following the sudden decrease in radius. We consider it possible that,
when mass transfer starts again, a phase of ``radio--ejection'' begins 
(\cite[Burderi \etal\ 2001]{burderi2001}; \cite[Burderi \etal\ 2002]{burderi2002}), in which mass accretion onto the NS is no longer allowed because of the pressure from the radio pulsar wind. In this case,
the matter is lost from the system at the inner lagrangian point, carrying away
angular momentum and altering the period evolution. This will occur for magnetic 
moments of the NS in a range $\sim 2 - 4 \times 10^{26}\, {\rm G}\,{\rm cm}^3$,
which is the most populated range for binary MSPs.

Turning now to the evolution below \Pbif,
it is well known that, if the secular evolution 
of LMXBs is similar to that of cataclysmic variables (CVs), one should expect
many systems at $P\simlt 2\,$hr, and  
a minimum orbital period similar to that of CVs, namely $\sim 80\,$min.
On the contrary, there are very few of these systems, and instead several ``ultrashort''
period binaries. In particular, three LMXBs in the field (which are also X-ray MSPs) and
one in a GC (X1832--330 in NGC~6652) are concentrated near
 \Porb $\sim 40\,$min. In addition there are two other ultrashort
period systems in GCs. While for GCs we may think that these systems
were formed by capture of a white dwarf by the NS, the field systems 
should have arrived at this period by secular evolution. Models have been constructed
by Nelson \& Rappaport (2003) and \cite{podsi2002}, and all imply that the donors
began Roche lobe overflow at periods just slightly below \Pbif,
so that the donor evolved to become a degenerate dwarf
predominantly composed of helium, but having a residual hydrogen abundance $<$10\%.
Until hydrogen is present in the core of the donor star, in fact, the evolution 
proceeds towards short \Porb (convergent systems). If the hydrogen content
left is very small, the mass radius relation when these objects become degenerate
is intermediate between that of hydrogen-dominated brown dwarfs and that 
of helium white dwarfs, so that smaller radii and 
shorter \Porb\ will be reached before radius and period increase again.

One problem of this scenario is that {\it there is a very small interval of
initial \Porb \/} which allows this very peculiar evolution: in most cases, either
a helium core is already formed before the mass transfer starts, and the system
evolves towards long \Porb, or there is enough hydrogen that the
system is convergent, but the minimum period is similar to that of CVs, and cannot
reach the ultrashort domain.
\cite{podsi2002} notice that, for a $1\,M_{\odot}$ secondary, the initial period range 
that leads to the formation of ultracompact systems is 13--18~hr. 
Since systems that start mass transfer in this period range might be naturally 
produced as a result of tidal capture, this could perhaps explain the large
fraction of ultracompact LMXBs observed in GCs. However,
quantitatively, this appears highly unlikely (van der Sluys et al.\ 2005).

In any case, this does not
apply to LMXBs in the field. In her PhD thesis, A.\ Teodorescu (2005) derived
the period distribution expected for LMXBs from convergent systems
under several hypotheses, and compared it with the available observed 
period distribution. An expected result was that the range \Porb$<2\,$hr is very populated
and the distribution is inconsistent with observations, unless we can suppress the 
secular evolution of all the systems below the ``period gap,'' which should occur
at about the same location as in CVs. One can consider several different possibilities
to do this: 

\noindent
(1) The lack of \Porb's $<$2~hr is again a consequence of radio--ejection: 
after the period
gap is traversed by the detached system, when the mass transfer resumes, it
is prevented by the pulsar wind pressure, the matter escapes from the system with high
specific angular momentum, and the evolution is accelerated. Indeed, this is
probably occurring in the system containing pulsar~W in 47~Tucanae, which has
 \Porb=3.2~hr. This system exhibits
X-ray variability which can be explained by the
presence of a relativistic shock within the binary that is regularly
eclipsed by the secondary star (Bogdanov \etal\ 2005). 
The shock can then be produced by the interaction
of the pulsar wind with a stream of gas from the companion
passing through the inner Lagrange point (L1), a typical case of what is expected
in radio--ejection (Burderi \etal\ 2001).
This mechanism could affect all the systems which enter a period gap. Notice that
only systems which end up at ultrashort periods {\it do not\/} detach during the
secular evolution, and they only might have a
``normal'' secular evolution. Thus both the lack of systems at \Porb$<2\,$hr and the
presence of ultrashort periods could be due to this effect.

\noindent
(2) ``Evaporation'' of the donor, due to the the pulsar wind impinging on,
and ablating material from, the surface of the companion (Ruderman \etal\ 1989)
is another possible mechanism, with results not so different from the previous case.

\noindent
(3) It is possible that the secular evolution almost never begins when the donor is
not significantly evolved. This can be true only if binaries are mostly 
formed by tidal capture, in which the NS captures a main-sequence star only at
separations $\simlt 3\, R_*$ (Fabian \etal\ 1975). This might happen in 
GCs, but we need to explain the \Porb\ distribution of {\it all\/} the LMXBs in the 
Galaxy. We could then reconsider the possibility that {\it most\/} of the 
field LMXBs were in fact formed in GCs, which were later destroyed (e.g.,
by tidal interactions with the Galactic bulge; but see Sec.~11). 

There are other specific cases that we must take into account when discussing
evolutions starting close to \Pbif. The famous interacting MSP binary
in NGC~6397, PSR J1740--5340 is such a case. At \Porb=35.5\,hr, it is in a
radio--ejection phase and the companion has certainly not 
been captured recently in a stellar encounter: it is an evolved subgiant, as
predicted by the secular evolution models (Burderi \etal\ 2002), 
and as confirmed by the CN cycled chemistry of the donor envelope, 
observed by Sabbi \etal\ (2003b) and predicted by Ergma \& Sarna (2003). 
We suspect that PSR J1748--2446ad in Terzan~5
is also in a radio--ejection phase (Burderi \etal\ 2006), but the
lack of information on the donor precludes a very secure interpretation. 
At \Porb=26.3\,hr, again, the donor should be in an early subgiant stage, and
have evolved very close to \Pbif. 

Finally, the whole period distribution of binary MSPs in GCs is 
consistent with a very high
probability of the onset of mass transfer being close to \Pbif. In fact,
there is a large group having \Porb\ from 0.1 to 1 day, a range not covered at 
all by the ``standard'' evolutions in \cite{podsi2002}, but which results easily from
the range of initial periods between those leading to ultrashort period binaries
and those above the bifurcation (Teodorescu 2005). 
In addition, there are several binary MSPs in GCs
for which the white dwarf mass is very low ($0.18 - 0.20\,M_{\odot}$), close to the minimum mass
which can be formed by binary evolution (Burderi \etal\ 2002), indicating again
evolution starting at a period slightly larger than \Pbif.

\section{Population Synthesis with Dynamics}

Ivanova and collaborators have developed a new simulation code to study the formation and 
retention of NSs in clusters, as well as the formation
and evolution of all compact binaries in GCs. This code is described in 
Ivanova \etal\ (2005, 2006).
The method combines the binary population synthesis code {\tt StarTrack} (Belczynski \etal\ 2002,
2007)
and the {\tt Fewbody} integrator for dynamical encounters (Fregeau \etal\ 2004).
Compared to other numerical methods employed to study dense stellar systems, this method 
can deal with very large systems, up to several million stars, and with 
large fractions of primordial binaries, up to 100\%,
although the dynamical evolution of the cluster is not treated in a fully self-consistent manner.

In addition to the formation of NSs via core collapse, these simulations take into account
NSs formed via electron-capture supernovae (ECS).
When a degenerate ONeMg core reaches a mass $M_{\rm ecs}=1.38\,M_\odot$,
its collapse is triggered by electron capture on $^{24}$Mg and $^{20}$Ne
before neon and subsequent burnings start and, therefore, before the
formation of an iron core (see, e.g., Nomoto 1984).
The explosion energy of such an event is significantly lower than that inferred 
for core-collapse supernovae (Dessart \etal\ 2006),
and therefore the associated natal kick velocities may be much lower.
There are several possible situations when a degenerate ONeMg core
can reach $M_{\rm ecs}$:

\begin{itemize}
\item  During the evolution of single stars: if the initial core mass is less than that required for neon ignition, $1.37\,M_\odot$,
the core becomes strongly degenerate. Through the continuing He shell burning, this core grows 
to $M_{\rm ecs}$.
The maximum initial mass of a single star of solar metallicity that leads to the formation
of such a core is $8.26\,M_\odot$,  and the minimum mass is $7.66\,M_\odot$.
This mass range becomes 6.3 to 6.9 $M_\odot$  for single stars with a lower GC metallicity
 $Z=0.001$.
The range of progenitor masses for which an ECS can occur depends also on the
mass transfer history of the star and therefore can be different in 
binary stars, making possible for more massive progenitors to collapse via ECS (Podsiadlowski \etal\ 2004).
\item As a result of accretion onto a degenerate ONeMg white dwarf (WD)
in a binary: accretion-induced collapse (AIC). In this case, a massive ONeMg WD steadily accumulates mass until it reaches
the critical mass $M_{\rm ecs}$.
\item When the total mass of coalescing WDs exceeds $M_{\rm ecs}$: merger-induced collapse (MIC).
The product of the merger, a fast rotating WD, can significantly exceed the
Chandrasekhar limit before the central density becomes high enough for
electron captures on $^{24}$Mg and $^{20}$Ne to occur, and therefore more massive NSs can 
be formed through this channel (Dessart \etal\ 2006).
\end{itemize}

Both metal-poor (Z=0.001) and metal-rich (Z=0.02) stellar populations have been studied
by Ivanova et al., who find that the production of NSs via core-collapse SNe (CC NSs) is 
20\% lower in the metal-rich population
than in the metal-poor population. In a typical cluster (with total
mass $2\times10^5\,M_\odot$, age $\sim10\,$Gyr, 1-dimensional velocity dispersion 
$\sigma=10$ km/s and central escape velocity 40 km/s)
about 3000 CC NSs can be produced, but less than 10 will typically be retained in the cluster.

ECS in single stars in a metal-rich population are produced from stars of higher masses,
but the mass range is the same as in metal-poor populations.
As a result, the number of ECS from  the population of single stars in the metal-rich case is 30\% smaller
than in the metal-poor population, in complete agreement with the adopted initial mass function (IMF).
The total number of NSs produced via this channel is several hundreds (and depends on the initial binary fraction),
but the number of retained NSs is higher than in the core-collapse case: about 150 NSs 
in a typical cluster.
The binarity smoothes the mass range where ECS could occur, and there are fewer differences between
the production of NSs via ECS in  binary populations of different metallicity.
The number of retained NSs produced via AIC and MIC is comparable to the number of ECS, about 100
in a typical metal-poor cluster.
Overall, one finds that, if a metal-rich GC has the same IMF and initial binary properties as a metal-poor GC, it will contain 30-40\% fewer NSs.

These simulations can also be used to examine the spatial distribution of pulsars and NSs
in clusters, although the present method only distinguishes between a central ``core''
(where all interactions are assumed to take place) and an outer ``halo.''
For a typical half-mass relaxation time $t_{\rm rh}=10^9\,$yr, about 50\% of all NSs
and 75\% of pulsars should be located in the core, and for a longer
$t_{\rm rh}=3\times 10^9$ yr, these fractions decrease to about 25\% and 50\% respectively.
Such predicted spatial distributions are in good agreement with observations of pulsars
in many GCs (Camilo \& Rasio 2005).

Ivanova et al.\ analyzed three main mechanisms for the formation of close binaries with NSs:
tidal captures, physical collisions with giants, and binary exchanges.
Very few primordial binaries with a NS can survive, except for those that
were formed via AIC. Typically $\sim 3\%$ of all NSs in a metal-poor GC can
form a binary via physical collision and $\sim 2\%$ via tidal captures, while
40\% of dynamically formed binary systems will start mass transfer (MT) in a Hubble time.
These number are slightly higher in the case of a metal-rich cluster,
and can be as much as two times higher in a cluster of the same metallicity but with 
a lower velocity dispersion, down to $\sigma=5$ km/s.
The binary exchange channel is more important for binary formation, as up to 50\%
of all NSs will be at some point members of binary systems, but only about 8\% of these
systems will start MT.

Overall, taking into account the formation rates of MT binaries with a NS and a MS star, 
and the duration of the MT phase, the probability that a cluster contains a NS--MS LMXB 
is almost unity, although most of them will
be in quiescence. For NS--WD binaries, the probability is $\sim 50\%$, but only a 
few percent of these will be in the bright
phase, when $L_{\rm X} > 10^{36}$ erg/s.
More LMXBs per NS are formed in metal-rich clusters, but since fewer NSs are produced 
and retained,
no significant difference in the resulting LMXB formation rate is found.

Finally, we note that if all ECS channels indeed work, too many NSs and pulsars 
(more than observed) are produced in these models.
Therefore, either one or more of the ECS channels (standard ECS, AIC or MIC) does not work,
or they have smaller allowed physical ranges where they can occur, or the kick associated 
with ECS could be larger.
Our current understanding of stellar evolution and NS formation and retention in GCs of 
different metallicities, coupled with the dynamical formation of mass-transferring 
binaries with NSs, cannot explain the
statistically significant overabundance of LMXBs in more metal-rich clusters.
Instead,  different  physics for the MT with different metallicities or different IMFs 
are required (Ivanova 2006).

\section{Green Bank Observations of Millisecond Pulsars in Clusters}

Since its first scientific observations five years ago, the Green Bank
Telescope (GBT), has uncovered at least 60 GC
pulsars, almost doubling the total number known.\footnote{There are at
  least 133 known GC pulsars, of which 129 are currently listed in
  P.~Freire's catalog at {\tt
    http://www2.naic.edu/$\sim$pfreire/GCpsr.html}.  For a recent
  review of GC pulsars, see \cite[Camilo \& Rasio (2005)]{cr05}.}
Almost all of these systems are MSPs, and the
majority are members of binaries. Incredibly, 30 of these new MSPs,
including many strange systems, are in the dense and massive bulge GC
Terzan~5 (with a total of 33), while another 10 are in the bulge
cluster M28 (for a total of 11).  Other clusters with new pulsars (and
the numbers new/total) are M30 (2/2), M62 (3/6), NGC~6440 (5/6), NGC~6441
(3/4), NGC~6522 (2/3), NGC~6544 (1/2), and NGC~6624 (4/5).

Most of the GCs with new pulsars are in the Galactic bulge, with large
columns of ionized gas along the lines of sight.  Almost all of the
new pulsars have been found using wide bandwidth (600\,MHz)
observations centered near 2\,GHz, a relatively high radio frequency
for pulsar searches.  Such observations are much less affected by
interstellar dispersion and scattering than traditional searches (at
1.4\,GHz or 430\,MHz), resulting in greatly improved search
sensitivities, particularly for the fastest MSPs.

Pulsar timing solutions using the GBT now exist for almost 50 of the
new MSPs, as well as for an interesting binary MSP found with the GMRT
\cite[(NGC~1851A; Freire \etal\ 2004)]{fgri04}.  These timing solutions
provide precise spin and orbital parameters, which are useful for
probing many aspects of NS physics, binary evolution,
and cluster dynamics.  In addition, the highly precise astrometric
positions (with typical errors of $\lesssim$0.1\arcsec) allow
additional probes of cluster dynamics and possible identification of
pulsar companions at optical or X-ray wavelengths.

Using the ensemble of 32 Terzan~5 (Ter~5) MSP timing solutions, the positions
of the pulsars with respect to the cluster center allow a statistical
measurement of the average NS mass ($\simeq$1.35$-$1.4\,$M_{\odot}$).
Similarly, the pulsar positions and dispersion measures (DMs; the
integrated electron column density along the line of sight to the
pulsar) provide a unique probe of interstellar medium electron density
variations over 0.2$-$2\,pc scales and show that they are not
inconsistent with Kolmogorov turbulence.  Several of the brighter
Ter~5 pulsars with timing solutions encompassing older Parkes
observations are beginning to show evidence for proper motions.
Average proper motion values from GC MSPs may provide the best proper
motion measurements of highly reddened clusters like Ter~5.
Finally, comparisons of the spin-period and luminosity distributions
of the 33 pulsars in Ter~5 and 22 in 47~Tuc show that they are
significantly different, and hence may be related to the properties
and dynamics of the GCs.

Among the interesting new pulsars are Ter~5E, a 2.2\,ms pulsar in a
60\,day orbital period (the 2$^{\rm nd}$ longest of any cluster MSP,
the longest is in the low-density cluster M53); Ter~5N, an 8\,ms pulsar with
a likely CO white dwarf companion, the first known in a GC; five
``black-widow''-like systems (M62E, Ter~5O, Ter~5ae, M28G, and M28J)
with few-hour circular orbits and $\sim$10$-$40\,$M_{\rm Jupiter}$
companions; and at least seven eclipsing binaries (M30A, Ter~5O, Ter~5P,
Ter~5ad, NGC~6440D, NGC~6624F, and M28H).

Several of the above systems hint at production mechanisms involving
stellar interactions.  But there are two other classes of very
interesting pulsars that are almost certainly produced via exchange
interactions: pulsar$-$``main-sequence'' binaries, and highly
eccentric ($e > 0.25$) binaries.  Recent 2-GHz GBT searches have
uncovered at least two of the former, and (amazingly) nine of the
latter.

Ter~5ad is the fastest MSP known \cite[($P$=1.396\,ms; Hessels \etal\
2006)]{hrs+06} and finally beats the 23-yr-old ``speed'' record
established by the first MSP discovered \cite{bkh+82}.  Ter~5P is the
5$^{\rm th}$ fastest MSP known.  Both systems are in circular binaries
($P_{\rm orb}$=26\,hr for Ter~5ad and 8.7\,hr for Ter~5P) with companions
of mass $\gtrsim$0.14\,$M_{\odot}$ for Ter~5ad and
$\gtrsim$0.36\,$M_{\odot}$ for Ter~5P.  Both systems are eclipsed for
$\sim$40\% of their orbit, yet on some occasions the eclipses appear
to be irregular (of different duration or possibly of variable depth).
These systems appear to be very similar to the fascinating MSP
J1740$-$5340 in NGC~6397 (\cite[D'Amico \etal\ 2001b]{dpm+01}).

Timing solutions for Ter~5ad and~P associate both pulsars with hard
X-ray point sources detected in a {\it Chandra\/} observation of
the cluster.  In addition, both systems exhibit extremely large orbital
period derivatives ($\dot P_{\rm orb}$$\sim$7$\times$10$^{-9}$) and
numerous (4 or more) higher-order period derivatives, likely due to
tidal interactions with the companion stars. Upcoming {\it HST\/} ACS
and near-IR {\it VLT\/} adaptive-optics observations may identify
``bloated'' companions, as for PSR J1740$-$5340 \cite{fpd+01}.

These two systems raise many questions: Why has it taken so long to
find a new ``fastest MSP''?  Do faster systems exist?  Why does
Ter~5 have 5 of the 10 fastest MSPs known in the Galaxy and {\it
  the\/} 5 fastest-spinning pulsars known in the GC system?  Can the
large orbital period variations constrain tidal circularization
theory?  Is the X-ray emission from magnetospheric pulsations, an
intra-binary shock, or some combination of both?

The second class of exchange products are the highly eccentric
binaries.  M15C, a double NS system, was the first highly-eccentric
binary discovered in a GC \cite{agk+90}, but it took ten years to find
the next, NGC~6441A \cite{pdm+01}.  Soon afterwards, M30B \cite[(found
with the GBT; Ransom \etal\ 2004)]{rsb+04} and then NGC~1851A
\cite[(currently being timed with the GBT; Freire \etal\
2004)]{fgri04} were detected.  The recent GBT 2-GHz surveys, though,
have uncovered {\it nine} additional highly-eccentric binaries: 6 in
Terzan5 (I, J, Q, U, X, and Z), 2 in M28 (C and D), and 1 in NGC~6440
(B).

Eccentric MSP binaries systems can be important probes of NS physics,
as they provide a way to constrain (or even directly measure) the
masses of fully-recycled pulsars.  Given the angular reference that an
ellipse provides, pulsar timing can easily measure the orbital advance
of periastron.  If the companion star is compact, the advance is
dominated by general relativistic effects and determines the total
system mass ($M_{\rm tot}$). The amount of mass required to spin-up a
MSP is currently unknown: the double NS systems with
precisely determined masses are only partially recycled, and there are
only a handful of mass measurements for fully recycled pulsars
\cite[(Stairs \etal\ 2004; Lattimer \& Prakash 2004)]{sta04}.  Since
the recycling scenario in general creates binary MSPs in circular
orbits (due to tidal circularization during the accretion phase),
these systems are {\it only\/} produced during interactions in dense
stellar systems (Rasio \& Heggie 1995).

Timing solutions from the GBT are now available for all of the known
highly eccentric binaries except for M15C \cite[(although see Jacoby
\etal\ 2006)]{jcj+06}, M30B (which has only been detected once, likely
due to strong scintillation effects), and Ter~5U (a very strange system
with $P_{\rm orb}$=1.8\,d, $e$$\simeq$0.27, and a minimum companion mass
of only 25\,$M_{\rm Jupiter}$).  From these 10 timing solutions, the
advance of periastron is highly significant in 9, indicating total
system masses between 1.6$-$2.5\,$M_{\odot}$.  Such values are
expected for recycled pulsars (with the NS mass being
1.4$-$2\,$M_{\odot}$) with white-dwarf-like companions, indicating that the
periastron advance is likely dominated by general relativity and not by
classical effects.

Two of these systems (Ter~5I and~J) appear to have ``massive'' NSs
($\sim$1.7\,$M_{\odot}$), which constrain the equation of state (EOS)
of matter at nuclear densities, possibly ruling out very soft EOSs or
those with strange-matter components \cite{rhs+05}.  Over
the next couple of years, measurements of the relativistic $\gamma$ parameter for
Ter~5I and possibly of the Shapiro delay for M28C are likely.  These
measurements, if the companions are white dwarfs, will provide accurate masses
for the NSs.  Several of the other eccentric binaries are interesting
as well: Does Ter~5Q (with $P$=2.8\,ms and $P_{\rm orb}$=30\,d) have a
NS companion?  How do you create a highly eccentric binary like Ter~5U
with a 25$-$30\,$M_{\rm Jupiter}$ companion?  Why was Ter~5Z not
ejected from the core when the interaction that made it eccentric
occurred?  Why does M28D (with $P$=79.8\,ms and
$P_{\rm orb}$=30\,d) appear to be a ``young'' pulsar (characteristic
age $t_c$$\sim$1$\times$10$^7$\,yr)?  Was it really formed (and
possibly partially-recycled) only recently?

Timing observations are ongoing for most of the new pulsars mentioned
here, and will continue to refine known parameters, to allow searches for
planetary companions, to measure secular effects possibly due to unseen
companions or stellar encounters, and likely to determine the proper
motions of the clusters.  In addition, searches of all timing data and
GBT observations of other clusters are underway.  We fully
expect the GBT to uncover many more GC MSPs, including some new
surprises, in the coming years.

\section{Parkes Observations of Radio Pulsars in Clusters}

The Parkes Globular Cluster Pulsar Search is a project started in the
mid 1990s as a side search project of the Parkes Multibeam Pulsar
Survey. Observations with the Parkes radio telescope have been
extensively performed with the central beam of the multibeam receiver
and the collected data have been processed with dedicated
algorithms. This project has so far led to the discovery of 12 new MSPs
in 6 GCs for which associated pulsars were
previously unknown. This section summarizes some recent results obtained from
timing these sources.

\subsection{NGC 6266}

The cluster NGC~6266 hosts six MSPs. The
first three of them, PSR J1701--3006A, B and~C, have been discovered in the 
framework of the Parkes
Globular Cluster Pulsar Search by \cite{dlm+01}, while the other
three have been discovered with further GBT observations by
\cite{jcb+02}.

All six pulsars in this cluster are members of binary systems
(\cite[D'Amico \etal\ 2001a]{dlm+01}; \cite[Possenti \etal\ 2003]{pdm+03}; \cite[Jacoby \etal\ 2002]{jcb+02}). This unusual occurrence is very unlikely
to be due to chance. It is possible that the absence of isolated
pulsars in this cluster is related to the peculiar dynamical state of
the cluster that lowers the rate at which binaries are disrupted via
dynamical encounters.

A peculiar object in this cluster is the MSP
PSR J1701--3006B (\cite[Possenti \etal\ 2003]{pdm+03}). It is in the emerging family
of eclipsing MSPs with relatively massive
companions. Several observations of this pulsars as it transits at the
superior conjunction show distortions in the signal that can be
ascribed to the pulsar motion through the companion wind matter. The
ablation timescale for the companion is compatible with the
pulsar's age only if less than 10\% of the released mass is ionized,
which appears unlikely. The alternate possibility is that mass loss is
due to nuclear evolution of the companion, analogous to the case of
the eclipsing system in NGC~6397 (\cite[Possenti \etal\ 2003]{pdm+03}).

\subsection{NGC 6397}

PSR J1740--5340 is a peculiar MSP in
NGC~6397 (D'Amico \etal\ 2001a,b). It is a binary source, whose
radio signal suffers eclipses as the pulsar approaches superior
conjunction (\cite[D'Amico \etal\ 2001b]{dpm+01}). The
extent of the eclipses strongly depends on the observing frequency. At
1.4\,GHz eclipses last up to 40\% of the orbit, and signal distortion
is observed at all orbital phases. Such distortions become less
prominent at an observing frequency of 2.3\,GHz and become nearly
absent at 3.0\,GHz.

Optical observations of the companion have revealed several
phenomena that occur as the pulsar emission interacts with the
companion surface:

\begin{itemize}
\item[i] The study of H$_{\alpha}$ lines indicates that matter is
swept away in a cometary tail by a radio ejection mechanism
(\cite[Sabbi \etal\ 2003a]{sgf+03});
\item[ii] The presence of He lines in absorption may indicate that
hot barbecue-like strips on the companion surface are heated by a
highly anisotropic pulsar flux (\cite[Sabbi \etal\ 2003b]{sgf+03});
\item[iii] The signature of an enhanced lithium abundance on the
companion surface may be perhaps ascribed to lithium production from
nuclear reactions triggered by accelerated particles flowing from the
pulsar (\cite[Sabbi \etal\ 2003b]{sgb+03}).
\end{itemize}

The strong distortion suffered by the radio signal from the pulsar
makes it very difficult to obtain a fully coherent timing solution for the
pulses. This is illustrated by a recent optical
determination of the companion position (\cite[Bassa \& Stappers 2004]{bs04}), which is
inconsistent with the previous determination obtained from the pulsar
timing (\cite[D'Amico \etal\ 2001b]{dpm+01}). The measurement of the companion position by
\cite{bs04} allowed the determination of a new timing solution
(\cite[Possenti \etal\ 2005]{pdc+05}), which will be updated with timing observations at
3.0\,GHz.

\subsection{NGC 6441}

The cluster NGC~6441 hosts PSR J1750-3703 (\cite[D'Amico \etal\ 2001a]{dlm+01}), a
binary pulsar in a highly eccentric orbit ($e=0.712$) with a
relatively massive companion. Since a data span of about five years is
now available, the periastron advance for this system is actually
measured with a precision of about 20$\sigma$ (\cite[Possenti \etal\ 2006]{pcc+06}), 
which gives a total mass for this binary $M_{\rm
tot}=2.20\pm0.17\,M_{\odot}$. This result can be combined with the
mass function for this system and the minimum measured mass for a NS
to obtain a range for the mass of the companion
$0.6\,M_{\odot}\,\leq\,M_{\rm C}\,\leq\,1.17\,M_{\odot}$
(\cite[Possenti \etal\ 2006]{pcc+06}). This makes it unlikely that  the companion is 
another NS.

\subsection{NGC 6752}

NGC~6752 contains five known MSPs.
The discovery of the binary pulsar PSR J1910--5959A
(\cite[D'Amico \etal\ 2001a]{dlm+01}) allowed the subsequent discovery of four more isolated
pulsars (\cite[D'Amico \etal\ 2002]{dpf+02}).

All five pulsars in this cluster show peculiar
features. PSR J1910--5959B and~E, located within a few
arcseconds from the cluster center, show a large negative value for the
spin period derivative (\cite[D'Amico \etal\ 2002]{dpf+02}). These negative values are
ascribed to the motion of these objects inside the cluster potential
well. PSR J1910--5959D is also located close to the cluster core
(\cite[D'Amico \etal\ 2002]{dpf+02}). Its spin period derivative is positive and of
the same order of magnitude as the values for PSR J1910--5959B and
PSR J1910--5959E, implying that, for this pulsar as well, the spin period
derivative is affected by the cluster potential
(\cite[D'Amico \etal\ 2002]{dpf+02}). These measurements allow us to investigate the mass-to-light
ratio in the central region of the cluster. A lower limit
$M/L_{V}\,\geq\,5.5\,M_\odot/L_\odot$ has been obtained
by \cite{fps+03}. Such a high value indicates the presence of a large
number of low-luminosity objects in the cluster core.

PSR J1910--5959A and~C are located in the outskirts
of the cluster, namely $\theta_{\perp, {\rm PSR\,A}}\,=\,6.3'$ and
$\theta_{\perp, {\rm PSR\,C}}\,=\,2.7'$ (\cite[D'Amico \etal\ 2001a]{dlm+01}; \cite[D'Amico \etal\ 2002]{dpf+02}; \cite[Corongiu \etal\ 2006]{cpl+06}),
values that put these two objects in first and second place,
respectively, among GC pulsars that show a large
offset from the cluster center. These unusual positions have been
 investigated in detail by \cite{cpg02} and \cite{cmp03}. The most probable explanation
invokes the ejection of these objects from the cluster core by
dynamical interactions with a central massive object that may be
either a single massive black hole or a binary black hole of
intermediate mass (\cite[Colpi \etal\ 2003]{cmp03}).

The recent measurement of proper motions for PSR J1910--5959A and
PSR J1910--5959C (\cite[Corongiu \etal\ 2006]{cpl+06}) shows that they are compatible with each
other, but they are not in agreement with the proper motion of the cluster as
determined from optical observations. Further observations of this
cluster will soon allow us to determine the proper motion of the
pulsars in the cluster core (certainly belonging to the cluster) and
the comparison between these proper motions and those of pulsars~A and~C will 
then establish whether the more distant pulsars are truly associated with the cluster.

\section{Chandra Observations of X-ray Sources in Clusters}\label{heinke}

The {\it Chandra\/} X-ray Observatory has provided fundamental new insights into the nature of faint ($L_X\sim10^{30-34}$ ergs/s) GC X-ray sources, through its superb spatial resolution and moderate spectral resolution. See Verbunt \& Lewin (2006) for a fuller (but dated) review, and the next section for some {\em XMM-Newton\/} results.  
The best-studied cluster (and the main focus in this section) is 47 Tucanae (47~Tuc), which has been well observed with \Chandra (detecting 300 X-ray sources), \HST in the optical and UV, and Parkes for pulsar timing observations (detecting 22 pulsars; see Sec.~7).  

\subsection{Low-mass X-ray Binaries}

The bright sources in GCs have long been known to be accreting neutron stars; X-ray bursts have now been detected from all Galactic GCs hosting luminous X-ray sources (in't Zand \etal\ 2003).  
{\it Chandra}'s resolution resolved a longstanding puzzle about the M15 LMXB: the optically identified LMXB in M15 is seen edge-on, not allowing direct view of the accreting (likely) neutron star; and yet X-ray bursts from a neutron star surface have been seen from M15.  
This puzzle is resolved by the identification of a second LMXB in M15 (White \& Angelini 2001).  This second LMXB has a period of 22.6 minutes (Dieball \etal\ 2005); this is the third neutron star accreting from a white 
dwarf (an ultracompact XRB; see Sec.\ 4) known in a GC.   

\subsection{Transient LMXBs}

\Chandra\ has allowed the identification of the quiescent counterparts to three transient LMXBs.  Quiescent LMXBs, at $L_X\sim10^{32}-10^{34}$ ergs/s, tend to show soft spectra, dominated by a $\sim$0.3 keV blackbody-like spectrum.  This component can be fit with a neutron star hydrogen atmosphere model, with implied radius of 10-15 km (Rutledge \etal\ 2002a).  This radiation is commonly thought to be produced by heating of the core during accretion, which will slowly leak out over $10^4$ years (Brown \etal\ 1998). A second harder component (of unknown origin) is often required above 2 keV, typically fit by a power-law with photon index 1-2.
Two of the transient LMXBs observed in quiescence fit this model; those in NGC~6440 (in't Zand \etal\ 2001) and Terzan 1 (Cackett \etal\ 2006).  
In contrast, the spectrum of the transient in Terzan 5 requires only a power-law component, with a photon index of $1.8^{+0.5}_{-0.4}$, indicating that quiescent LMXBs may also have relatively hard spectra (Wijnands \etal\ 2005).

\subsection{Quiescent LMXBs}

In addition to the known quiescent counterparts of transient LMXBs, additional X-ray sources are seen in clusters with spectra and luminosities characteristic of quiescent LMXBs.  Spectral fitting of the brightest of these with neutron star atmosphere models gives inferred radii consistent with 10-12 km (Rutledge \etal\ 2002b).  Two such systems in the cluster 47 Tuc show regular eclipses at periods of 8.7 and 3.1 hours (Heinke \etal\ 2005a), and two systems have faint optical counterparts (Haggard \etal\ 2004).  Pooley \etal\ (2003) and Heinke \etal\ (2003b) showed that the numbers of quiescent LMXBs in different clusters scaled with the stellar interaction rate in those clusters, implying they are formed dynamically.

If the distance to the GC is reasonably well-known, then it is possible to constrain the radius (or a combination of radius and mass) of the glowing neutron star through fits to hydrogen atmosphere models.  This has the potential to improve our understanding of the composition of neutron star interiors, and thus the behavior of matter at high density.  In units of $R_{\infty}$(=$R*(1+z)$), constraints have been placed on the neutron star in $\omega$ Cen ($R_{\infty}=14.3\pm2.1$ km, Rutledge \etal\ 2002b), and on X7 in 47 Tuc ($R_{\infty}=18.3^{+3.8}_{-1.2}$ km, Heinke \etal\ 2006b); see below for XMM results.  Perhaps the largest remaining source of uncertainty in these calculations is the distance to the GCs; recent authoritative determinations of the distance to 47 Tuc by the subdwarf main-sequence fitting method and direct geometry give results which differ by 20\% (Gratton \etal\ 2003; McLaughlin \etal\ 2006).  

\subsection{Cataclysmic Variables}

Optical counterpart searches using \HST have identified 22 cataclysmic variables (CVs) in 47 Tuc through blue optical/UV colors and variability, eight of which have secure orbital periods (Edmonds \etal\ 2003a).  Ten CVs have also been identified in NGC~6752 (Pooley \etal\ 2002) and nine in NGC~6397 (Grindlay \etal\ 2001b).  These CVs have blue $U-V$ colors, but $V-I$ colors that are on or near the main sequence.  
This indicates that the secondaries dominate the optical light, which is in agreement with the identification of ellipsoidal variations in several of these systems. Comparison of these and other cluster CVs with Galactic CVs shows that cluster CVs have fainter accretion disks than Galactic CVs with similar periods (Edmonds \etal\ 2003b).  This suggests that cluster CVs have relatively low mass transfer rates.  However, the lack of dwarf nova outbursts from cluster CVs (Shara \etal\ 1996) may be an indication that cluster CVs tend to be strongly magnetic (e.g., Dobrotka \etal\ 2006).

CVs may be formed in GCs either dynamically or from primordial binaries.  Several \Chandra\ observational studies (Pooley \etal\ 2003, Heinke \etal\ 2003b, Pooley \& Hut 2006, Heinke \etal\ 2006b, Kong \etal\ 2006) as well as population synthesis studies (Ivanova \etal\ 2006) point to contributions by both mechanisms to the existing CV population in clusters.

\subsection{Active Binaries}

Numerous chromospherically active binaries (mostly close main-sequence binaries) have been identified in several GCs.  Sixty have been identified with \Chandra\ sources in 47 Tuc alone (Heinke \etal\ 2005b).  Bassa \etal\ (2004) and Kong \etal\ (2006) have recently shown that the population of active binaries in clusters, unlike CVs and LMXBs, is produced from primordial binaries; in the densest clusters (such as NGC 6397), these binaries have been largely destroyed (Cool \& Bolton 2002).

\subsection{Radio Millisecond Pulsars}

MSPs have been detected in X-ray in several GCs (e.g., Bassa \etal\ 2004).  
The deep observations of 47 Tuc have detected all 19 MSPs with known positions (Bogdanov \etal\ 2006), showing that in most cases their X-ray spectra are dominated by thermal emission from their hot polar caps.  
Comparison of the X-ray spectra of unidentified sources in 47~Tuc with known MSPs and active binaries reveals that the majority of the unknown sources are active binaries, and constrains the total number of MSPs in 47 Tuc to $<$60, most likely $\leq30$ (Heinke \etal\ 2005b).  This helps to resolve a suggested discrepancy between the birthrates of LMXBs and MSPs in Galactic GCs. 

A few of the MSPs in 47 Tuc (and elsewhere) show harder X-ray spectra, suggestive of nonthermal synchrotron or shock emission.  
One of these, 47 Tuc-W, shows long X-ray eclipses, indicating the X-rays are produced in a shock near the companion from matter that continues to overflow the companion's Roche lobe (Bogdanov \etal\ 2005)--making this a ``missing link'' between LMXBs and MSPs.  
These discoveries have greatly improved our understanding of the evolution of LMXBs into MSPs in clusters.

\section{XMM-Newton Observations of X-ray Sources in Clusters}

Observations of Galactic GC faint X-ray sources made
with the two X-ray satellites {\em XMM-Newton\/} (e.g., Webb \etal\ 2006; Webb \etal\ 2004; Gendre \etal\ 2003a,b; Webb \etal\ 2002) and {\em Chandra\/} (Heinke \etal\ 2006; Pooley \etal\ 2003; see Sec.~8) have revealed that 25 are neutron star XRBs.  {\em XMM-Newton\/} spectra of these systems are of
sufficiently high quality, even with only 30\,ks observations, to well
constrain the mass and radius of the neutron star, using neutron star atmosphere models (e.g., Zavlin \etal\ 1996; Heinke \etal\ 2006)  and taking advantage of
the fact that their distances and interstellar absorptions are well
constrained due to their situation in a GC (Servillat \etal\ in preparation;
 Gendre \etal\ 2003a,b).  The masses and radii
are essential for constraining the (poorly known) equation of state of the
nuclear matter in these very compact stars.

Gendre \etal\ (2003b), Pooley \etal\ (2003) and Heinke \etal\ (2003) have also used the 
GC observations of faint X-ray
sources, coupled with the result that the bright X-ray sources ($L_x >$
10$^{36}$\,erg\,s$^{-1}$; Hertz \& Grindlay 1983) are also neutron star
XRBs (see Verbunt \& Hut 1987 and references therein), to confirm, through observations, the theory
 that these objects are formed mainly through dynamical encounters.  This
implies a total population of approximately 100 neutron star XRBs
distributed throughout the 151 Galactic GCs
(Pooley \etal\ 2003).  This population is wholly insufficient to slow down the inevitable
core collapse of these self-gravitating stellar clusters if the energy
liberated by binaries interacting with
other cluster stars is indeed the internal energy source necessary to
counter the tendency of clusters to collapse (see Hut \etal\ 1992 for a review).  

Cataclysmic variables (CVs) exist in much greater numbers in GCs.
Indeed Di Stefano \& Rappaport (1994) predict of the order
one hundred CVs in a single Galactic GC.  This prediction is born
out by observations, for example more than 30 CVs have  been detected
in 47~Tuc using X-ray observations (Heinke \etal\ 2005) and
approximately 60 candidate CVs have been identified in NGC~2808 using
UV observations (Dieball \etal\ 2005).  Thus, although we do
not yet know the whole population size of CVs in the GCs
observed (unlike for the brighter, soft neutron stars) with which to determine their
formation mechanisms and numbers, it is apparent  that they exist in
large numbers and thus it is possible that they are important to the
cluster's fate.

With more and more cataclysmic
variables identified in Galactic GCs, one striking and unexplained
difference has become clear between cluster CVs
and field CVs. Cluster CVs show a distinct lack of
outbursts (characterized by a steep rise in the flux by several orders
of magnitude) compared to field CVs.  Due to the proximity of the
white dwarf and its companion in a CV, material is
accreted from the companion star and stored  in the accretion disk
around the white dwarf whilst it loses sufficient angular momentum to
fall onto the compact object.  Outbursts are believed to occur when
too much material builds up in the disc, increasing both the density
and the temperature, until the hydrogen ionizes and the viscosity
increases sufficiently for the material to fall onto the white dwarf
(Osaki 1974; Meyer \& Meyer-Hofmeister 1981; Bath \& Pringle 1981).  
Many types of field CVs show such outbursts every few weeks to
months.  However, only very few GC outbursts have been
observed (e.g., Paresce \& de Marchi 1994; Shara \etal\ 1996, 1987) and
it is unclear why this should be.

It was originally  suggested that cluster CVs may be mainly
magnetic (see the five CVs in Grindlay 1999).  Magnetic CVs
have accretion discs that are either partially or totally disrupted by
the strong white dwarf magnetic fields and these two types are known as intermediate polars
and polars, respectively.  Material is channelled along the field lines
onto the white dwarf, although in the case of intermediate polars, a
truncated disk can exist and these systems can undergo a limited
number of outbursts (e.g., Norton \& Watson 1989).  Recently
it has been proposed that it may not simply be the magnetic field that
is responsible for the lack of outbursts. Dobrotka \etal\ (2006)
suggest that it may be due to a combination of low mass transfer rates
($\simlt 10^{14-15}$ g s$^{-1}$) and moderately strong white dwarf
magnetic moments ($\simgt 10^{30}$ G cm$^{3}$) which could
stabilize the CV discs in globular clusters and thus prevent most of them from
experiencing {\it frequent outbursts}.   This result suggests that  the
brightest globular cluster CVs in Ter~5s should be
intermediate polars. Ivanova \etal\ (2006) have also proposed that the
lack of outbursts is due to higher white dwarf masses (higher mean masses are observed amongst strongly magnetic isolated white dwarfs; Wickramasinghe \& Ferrario 2000).  
This is likely to be due to the difference in
the formation mechanisms of GC and field CVs, since a
substantial fraction of cluster CVs are likely to be formed through
encounters, rather than from their primordial binaries (Ivanova \etal\ 2006).

Intermediate polars show modulation on the spin period
(typically $\sim10^2 - 10^3\,$s) of the accreting white dwarf
which can be detected through Fourier analysis. For example,
 Parker et al.\ (2005) showed that 70\% of the
intermediate polars that were observed with {\em ASCA\/} and {\em RXTE\/}
showed this modulation. Thanks to the sensitivity of the {\em
XMM-Newton\/} satellite, observations made with this observatory of the
cluster NGC~2808 revealed that the brightest
CV in this cluster shows evidence for a modulation with a 430\,s period
(Servillat \etal\, in preparation).  This is likely to be the
modulation on the spin period, supporting an intermediate polar
identification. Low-resolution spectra of the brightest
CV (candidate) in the cluster M22 (Webb \etal\ 2004; Webb
\etal\, in preparation) also show some evidence for the He 4686\AA\ line in
emission, indicative of a magnetic white dwarf (e.g., Szkody \etal\ 2005).   As the CV has already been observed to  outburst
(Anderson \etal\ 2003; Bond \etal\ 2005; Pietrukowicz \etal\ 2006), it would indicate that 
this source is also an intermediate polar,
again supporting the idea that cluster CVs have moderate
magnetic field strengths, in part responsible for their lack of
outbursts.

We now turn briefly to possible formation mechanisms for cluster CVs.
It is now believed that there are two populations of CVs in GCs,
those formed dynamically (as the neutron star LMXBs), 
thought to be located in the dense cluster cores, and those
that have evolved from a primordial binary without undergoing any
significant encounter.  This latter population may reside outside the
cluster core (Davies 1997), where the stellar density is much lower
than near the center. Naturally we expect that the more concentrated
GCs,  which have higher core densities, have higher
encounter rates, thus increasing the number of CVs formed through
encounters.  In addition, the timescales of encounters between
primordial binaries and single stars are shortened, thus decreasing
the number of primordial CVs.  The GCs that have been
observed with {\em XMM-Newton\/} are particularly well adapted to
searching for a primordial binary population, as they are low-density
clusters, chosen to ensure that we can resolve all the X-ray
sources (the angular resolution of the EPIC cameras is
approximately 6'' Full Width at Half Maximum of the Point Spread
Function).  In addition, {\em XMM-Newton}'s large collecting area
ensures that there are enough photons for a full spectral study of about
20\% of the sources detected, advantageous for identifying CVs using
X-ray data alone.  Several CVs have already been detected in the cores
of GCs, like AKO~9 in 47~Tuc
(Auri\`ere \etal\ 1989), which \cite{knig03} state was almost certainly formed
dynamically, either via tidal capture or in a three-body encounter.
Such dynamically formed CVs exist in other GCs, like
$\omega\,$Cen (e.g., Carson \etal\ 2000; Gendre \etal\ 2003a) and M22 (Webb \etal\ 2004).

Several X-ray sources in GCs studied
with {\em XMM-Newton\/} lie outside the half-mass radius and
have X-ray luminosities, spectra, colors, and lightcurves that indicate
they may be CVs.  Recently, Pietruckowicz \etal\ (2005) confirmed,
using optical photometry, that one of these X-ray sources (Webb \etal\ 2004) lying at 3.9
core radii from the centre of M22 is indeed a
CV.  It is possible that this CV was formed from a primordial
binary.  Ivanova \etal\ (2006) predict that as many as 37\% of the CVs
in a cluster like 47~Tuc should be formed from the
primordial binaries, thus one would expect an even greater percentage
for a lower-concentration cluster such as M22, supporting the
primordial formation mechanism.

\section{X-ray Luminosity Functions}

Populations of X-ray sources have been discovered with {\em Chandra\/} in all kinds of galaxies. These populations provide a novel approach to study the evolution of XRBs. This section summarizes recent results from the study of the X-ray Luminosity Functions (XLFs) in these extragalactic populations.

XLFs (in either differential or cumulative form) provide a useful tool for characterizing and comparing XRB populations. Cumulative XLFs are typically described by functional slope(s), breaks, and normalization. Each of these parameters is potentially related to the formation and evolution of XRBs in a given stellar population: the distribution of luminosities (slope) has been found to be related to the age of the population (see below); breaks in the XLF are a possible indication of multiple or evolving XRB populations in the same galaxy; the normalization is a measure of the total number of XRBs. Grimm \etal\ (2002) first reported differences in the XLFs of different types of binaries, by deriving the ``young-short-lived'' high-mass X-ray binary (HMXB) and ``old LMXB'' XLFs for the Milky Way. They found that the HMXB XLF is well fitted by a single power-law, while the LMXB XLF may show
 both high- and low-luminosity breaks. More recent studies, based on {\it XMM-Newton\/} and {\it Chandra\/} observations, are in general agreement with these early results, but also show a more complex reality (see review 
by Fabbiano 2006).

Early studies of the integrated X-ray luminosity of star-forming galaxies pointed to a tight connection between the number of XRBs and star formation activity (e.g., Fabbiano \etal\ 1988; Fabbiano \& Shapley 2002). More recently, comparisons of XLFs have suggested a dependence of the normalization on the star formation rate (SRF) of the galaxy (Kilgard \etal\ 2002; Zezas \& Fabbiano 2002; Grimm \etal\ 2003). Grimm \etal\ (2003), in particular, propose that all HMXB XLFs follow a similar cumulative slope of -0.6, and have normalization strictly proportional to the SFR.  
The XLFs of individual spiral galaxies (see Fabbiano 2006 and references therein) not always agree with this conclusion. However, deviations can be understood if the effect of XRB populations of different ages is considered. A particularly illuminating case is that of M83, a grand-design spiral with a nuclear starburst  (Soria \& Wu 2003). In M83, the XLF of the nuclear region is a power-law with a cumulative slope -0.7, reasonably consistent with Grimm \etal\ (2003); instead, the XLF of the outer disk is complex, suggesting several XRB populations. This XLF has a break (and becomes steeper) at luminosities above $8 \times 10^{37}\,$erg/s, suggesting an older XRB population than that of the nuclear region; below the break it follows a -0.6 power-law, but this power-law is interrupted by a dip. Soria \& Wu suggest that this complex disk XLF may result from the mixing of an older disk XRB population mixing with a younger (but aging) population of spiral-arm sources.

The best-studied example of an extreme young HMXB population so far is given by the deep {\em Chandra\/} study of Antennae galaxies (Fabbiano \etal\ 2003), which led to the discovery of 120 X-ray sources down to a limiting luminosity near $2 \times 10^{37}\,$erg/s (Zezas \etal\ 2006), and including 
about 14 ultra-luminous X-ray sources (ULXs). The cumulative XLF is well fitted with a single power-law of slope of $\sim -0.5$. There is a small deviation from this power-law near $2 \times 10^{38}\,$erg/s, suggesting a NS Eddington limit effect (however this result is not statistically significant; Zezas \etal\ 2007).

Two papers, reporting the analysis of samples of early-type galaxies, give a good picture of the LMXB XLF at luminosities greater than a few 10$^{37}\,$erg/s. Kim \& Fabbiano (2004) analyzed 14 E and S0 XLFs, corrected each individual XLF for incompleteness by means of simulations, and found that the corrected XLFs could be fitted (above $6 \times 10^{37}\,$erg/s) with similar steep power-laws.  Given this similarity, the data were co-added resulting in a significantly higher signal-to-noise XLF, which shows a break, formally at $4.5 \times 10^{38}\,$erg/s, and marginally consistent with the NS Eddington limit. Gilfanov (2004) using a sample of early-type galaxies and spiral bulges reached a similar conclusion. 

Gilfanov (2004) suggested that the global stellar mass of the galaxy or bulge is the driving factor for the normalization of the XLF (the total number of slowly-evolving LMXBs in a given stellar population). A somewhat different conclusion was reached by Kim \& Fabbiano (2004; see also Kim \etal\ 2006). They reported a correlation between the total LMXB luminosity of a galaxy and the K-band (stellar) luminosity, in agreement with the stellar-mass-normalization link, but the scatter of this correlation is large, and while independent of total K-band luminosity, is correlated with the specific frequency of GCs in the galaxies (a link previously suggested by White \etal\ 2002). The conclusion, similar to that suggested for Galactic LMXBs (Clark 1975), is that GCs have a special effect on the formation and evolution of LMXBs. These results, however, do not exclude the possibility that evolution of native field binaries is also important, a point stressed by Irwin (2005).

LMXBs as short-lived ultra-compact binaries formed in GCs, have been discussed by Bildsten \& Deloye (2004; see also Ivanova \etal\ 2005), who point out that their model can reproduce the observed XLF. Disruption of (or expulsion from) GCs could then give rise to LMXBs in the stellar field. Alternatively, LMXBs may form and evolve in the field (Verbunt \& van den Heuvel 1995; Kalogera 1998 and references therein). Field source evolution models (e.g., Piro \& Bildsten 2002) have not set constraints on the XLF, but predict that most high-luminosity LMXBs should be detached binaries with large unstable disk, and therefore recurrent transients. While more time monitoring observations are needed, transients are indeed detected in some cases (e.g., in NGC~5128; Kraft \etal\ 2001).

A large body of work has addressed the association of individual X-ray sources with GCs in E and S0 galaxies, and the properties of observed field and GC LMXBs (Sec.~11). A particularly relevant result from the recent paper by Kim \etal\ (2006), based on the analysis of the LMXB populations of 6 galaxies observed with {\it Chandra\/}, is worth mentioning here: no significant difference can be seen in the XLFs of LMXBs with and without a GC counterpart. Moreover, both XLFs extend to luminosities above $5 \times 10^{38}\,$erg/s, a regime where the accreting object is likely to be a black hole. Ivanova \& Kalogera (2006) have pointed out that high-luminosity LMXBs are likely to be field transients populating the XLF in outburst, and that the XLF can be considered as the footprint of the black-hole mass function, with a differential slope of -2.5 and upper mass cut-off at 
$20\,M_\odot$. However, the field and cluster LMXB XLFs are similar and both extend to high luminosities. 
Are black-hole X-ray binaries therefore present in GCs, despite their very low expected formation probability (Kalogera \etal\ 2004)?

Very recent work (Kim \etal\ 2006), based on deep {\em Chandra\/} observations of two nearby elliptical galaxies with well studied old stellar populations, NGC~3379 and NGC~4278, is addressing the low-luminosity LMXB XLF, at luminosities below a few times $10^{37}\,$erg/s, which are typical of the majority of LMXBs in the bulge of the Milky Way and M31. The LMXB XLF of the Milky Way (Grimm \etal\ 2002) becomes flatter at these lower luminosities. Gilfanov (2004) suggested a significant ``universal'' flattening below $5 \times 10^{37}\,$erg/s in the LMXB XLF. This flattening is suggested by a number of models (see, e.g., Bildsten \& Deloye 2004; Pfahl \etal\ 2003). Kim \etal\ (2006) demonstrate that there is no universal low-luminosity flattening of the LMXB XLF. The XLF of NGC~4278 is very well fitted with a continuous power-law with cumulative slope $-1$, down to $1 \times 10^{37}\,$erg/s. The XLF of NGC~3379 (extending down to near $10^{36}\,$erg/s) is also well represented by a similar power-law, but it presents a statistically marginal localized excess near $4 \times 10^{37}\,$erg/s. 

\section{Extragalactic Globular Cluster X-ray Sources}

Other galaxies, particularly ellipticals, have proved to be excellent
grounds for learning about what kind of GCs produce
bright X-ray sources.  These galaxies have GC systems up
to two orders of magnitude larger than the Milky Way's.  Furthermore,
they often show much more diversity than the Milky Way in terms of
metallicities and ages of the clusters.  On top of this all, there are
dozens of galaxies within 20 Mpc, so even if one galaxy fails to
provide a sufficiently large or diverse population of clusters for
studying a particular effect, one can co-add many galaxies.  Studies
of elliptical galaxies have been especially fruitful, since the
specific frequencies of GCs are larger in more massive,
later type galaxies.  Additionally, the GC samples are
better understood in elliptical galaxies than in spiral galaxies
because of the smoother field star backgrounds in elliptical galaxies
are easier to subtract off than the knotty emission in spiral galaxies.

It was determined early in the {\it Chandra\/} era that about half of
all X-ray sources in elliptical galaxies are in GCs (see,
e.g., Sarazin \etal\ 2001 and Angelini \etal\ 2001 for the first few
studies, and Kim \etal\ 2006 for an analysis of an ensemble of
galaxies).  This compares with matching fractions of 10\% in the Milky
Way (van Paradijs \etal\ 1995) or $\sim$20\% in M31 (Supper \etal\ 1997).  The fraction was found to increase continuously through
the Hubble tuning fork diagram from spirals to lenticulars to
ellipticals to cD galaxies (Maccarone \etal\ 2003).  It has
been suggested that a substantial fraction of non-GC
X-ray sources were originally formed in clusters (e.g., White \etal\ 2002), but recent work has shown both that the fraction of
X-ray sources in clusters increases with specific frequency
of GCs (Juett 2005), and that the ratio of X-ray to
optical luminosity increases more slowly than linearly with the
specific frequency (Irwin 2005).

One of the key areas of interest for extragalactic GC
studies is the determination of which cluster properties most affect
the likelihood that a cluster will contain an XRB.  The most
significant parameter is the cluster mass (Kundu \etal\ 2002), with
several studies finding that the probability a cluster will be an
X-ray source scales in a manner consistent with $M^{1.1\pm0.1}$ (Kundu
\etal\ 2003; Jordan \etal\ 2004; Smits \etal\ 2006).

The normalization in the number of XRBs per unit stellar
mass is still considerably larger than in galaxies' field populations,
where the number of LMXBs seems to be linearly
proportional to the stellar mass as well (Gilfanov 2004; Kim \&
Fabbiano 2004), so clearly the number of XRBs must be
related to the stellar interaction rates in the clusters.  Determining
the stellar interaction rate requires an understanding of the radial
profile of the cluster as well as its total mass, and this is rather
difficult to determine for most GCs at the 10-16 Mpc
distances of the best-studied elliptical galaxies.  Kundu \etal\ (2002) found for NGC~4472 that the half-light radius of a cluster
was a marginally significant parameter for predicting whether a
GC would have an X-ray source.  Jordan \etal\ (2004)
attempted to fit King models to the GCs in M87 and found
no statistically significant difference between the predictive power
of cluster mass and of inferred cluster collision rate for whether a
cluster would contain an X-ray binary.  There is only a weak
correlation between cluster core radius and cluster half-light radius,
except in the least concentrated clusters, and since the core radii of
nearly all Milky Way GCs are too small to be resolved,
even with the Hubble Space Telescope, at distances exceeding a few
Mpc.  It is thus not surprising that we cannot obtain any
quantitative information about the relation between cluster collision
rate and probability of containing an XRB by looking at Virgo
Cluster elliptical galaxies.

It was found by Smits \etal\ (2006) that the observed
$P({\rm LMXB})\propto$$M^{1.1}$ is consistent with bimodal pulsar kick-velocity 
distributions (e.g. Arzoumanian \etal\ 2003; Brisken \etal\ 2003), but not with a 
single Maxwellian kick-velocity distribution around 200 km/s (e.g., Hobbs \etal\ 2005).
Because binary evolution effects can produce low kick-velocity pulsars
(e.g., Pfahl \etal\ 2002; Dewi \etal\ 2005) it is not clear whether the
cluster XRB results have direct implications for the
controversies concerning isolated pulsar velocity distributions.

Aside from mass, the other key parameter which helps determine whether
a cluster will contain an X-ray source is its metallicity.  There were
some indications from the pre-{\em Chandra\/} era that this was the case,
based on the Milky Way and M31 (Grindlay 1993; Bellazzini \etal\ 1995), but the strong correlation in spiral galaxies between
metallicity and galactocentric radius, along with the relatively
strong tidal forces in the centers of spiral galaxies, left some doubt
about which was the underlying physical cause of the enhancement of
X-ray sources in metal-rich bulge GCs.  It has since
been proven clearly, in numerous elliptical galaxies, that metallicity
really is a strong predictor of whether a cluster will have
an X-ray source (Kundu \etal\ 2002; Di Stefano \etal\ 2003; Jordan \etal\ 2004; Minnitti \etal\ 2004; Xu \etal\ 2005; Posson-Brown \etal\ 2006; Chies-Santos \etal\ 2006).

Attempts have also been made to determine whether cluster ages affect
X-ray binary production, especially in light of theoretical
suggestions that there should be a peak in the X-ray source production
rate at ages of about 5 Gyr (Davies \& Hansen 1998).  In NGC~4365,
which has a substantial sub-population of intermediate-age 
clusters (Puzia \etal\ 2002; Larsen \etal\ 2003; Kundu \etal\ 2006),
it is clear that there is an effect of metallicity on the probability
a GC will be an X-ray source (Kundu \etal\ 2003);
roughly the same effect of metallicity is seen in NGC~3115 (Kundu \etal\ 2003), which has 
only old GCs (Puzia \etal\ 2002).
This argues in favor of the idea that the metallicity effect is
causual.  

Two viable possibilites have been suggested for this effect.  One is
irradiation-induced stellar winds (Maccarone \etal\ 2004), which
should be stronger in low-metallicity environments (Iben \etal\ 1997)
since the energy deposited by irradiation in a low metallicity star
cannot easily be dissipated by line cooling.  As a result, the metal-poor stars will lose much of their mass to the interstellar medium,
rather than to the accreting star, yielding effectively lower duty
cycles as bright sources.  The other model depends on the smaller
convection zones of metal-poor stars compared with metal-rich stars
(Ivanova 2006).  This leads to reduced cross sections for formation of
X-ray binaries through tidal capture, and to less efficient magnetic
braking, and hence lower accretion rates.  A distinguishing
characteristic of the models is that the irradiation wind model should
leave behind absorbing material which will leave an absorption
signature in X-ray spectra, while the convection zone model should
not.  Spectra of M31 clusters in the 0.1-2.4 keV ROSAT band are harder
in the more metal-poor clusters (Irwin \& Bregman 1999), while {\em Chandra\/}
spectra showed no correlation between X-ray spectrum and cluster
metallicity (Kim \etal\ 2006).  It is thus not clear whether the M31
results are a statistical fluke, or the {\em Chandra\/} data, with very little
sensitivity to X-rays below 0.7 keV, are simply not sensitive to this
effect. 

\section{Massive Black Holes in Globular Clusters}


There has been considerable debate as to whether evidence supports the
existence of massive central black holes in GCs. M15 has been the focus
for decades, and the latest observational results by van den Bosch \etal\ (2006) show 
no significance for a black hole
($1000\pm1000\,M_{\odot}$).  The same is true of 47~Tuc where McLaughlin \etal\ (2006) report $700\pm700\,M_{\odot}$. In both of these cases, the value
reported, while not significant, is consistent with that mass expected
from an extrapolation of the correlation between black hole mass and
host dispersion as reported in Gebhardt \etal\ (2000) and Tremaine \etal\ (2002). 

The case is different in G1, the largest cluster in M31.
In G1, Gebhardt, Rich \& Ho (2002) reported a mass of
$2\times10^4\,M_{\odot}$, which was subseqently challenged by Baumgardt \etal\ (2003), who argued against a central black hole. However, the latest
observations by Gebhardt \etal\ (2005) continue to argue for a
massive black hole using newer data. Whether GCs contain
black holes has significant effects on both the evolution of the
cluster and on how supermassive black holes grow. Thus it is very
important to understand possible number densities for these black
holes, and the current observational situation is not
satisfying. Theoretically, there are reasons to expect massive black holes in
clusters, although observations are required.

We now turn to very recent observations of $\omega\,$Cen.
This cluster is an ideal candidate to look for a central black hole. It
has one of the largest velocity dispersions among GCs,
implying it may have a large black hole mass. It is nearby both
allowing for any black hole influence to be well resolved and allowing
access to many stars used to trace the gravitational potential. With
an integrated velocity dispersion of 18~\kms, the expected black hole
is $10^4\,M_{\odot}$ which has a sphere of influence of 6\arcsec\ at the 4.8~kpc
distance o fthe cluster.  The issue with $\omega\,$Cen is that it may not be a globular
cluster, but has been suspected to be the nucleus of an accreted dwarf
galaxy (Freeman 1993; Meza \etal\ 2005). Thus, while a massive black hole 
in $\omega\,$Cen would not necessarily answer the question as to whether GCs
contain black holes, it would help establish the existence and
frequency of intermediate-mass black holes in general.

Kinematic data on the cluster come from two sources. Noyola, Gebhardt, \&
Bregmann (2007) used Gemini/GMOS-IFU data to measure the integrated
light in the central 3\arcsec\ and at 14\arcsec\ radius. Gebhardt \&
Kissler-Patig (2007) used individual velocities in the central
8\arcsec\ to measure the dispersion.  Both
observations are consistent, and, since one uses integrated light and
the other individual velocities, argue for a robust central dispersion
estimate.  The central velocity dispersion for the cluster is
$24\pm2$~\kms. The dispersion at 14\arcsec\ is 20~\kms. Beyond
25\arcsec, data have been compiled by van den Ven \etal\ (2006),
coming primarily from radial velocities of Xie \etal\ (2007). Noyola \etal\ and Gebhardt \& Kissler-Patig use orbit-based dynamical models and
require a central black hole mass of $4(\pm0.8)\times10^4\,M_{\odot}$.

The main arguments against having a black hole are allowing radial
orbital anisotropy and having a significant population of heavy
remnants. For the orbital anisotropy, there are two
considerations. Van de Ven \etal\ (2006) model $\omega\,$Cen using both
radial velocities and proper motions, at radii beyond 25\arcsec. They
find a distribution function very close to isotropic.  The amount of
radial anisotropy required to increase the central dispersion is
extreme (see Noyola et al.) and very inconsistent with the van de Ven \etal\ model. 
Furthermore, orbit-based models have been constructed which allow for any
orbital distribution consistent with the Jeans equations. For these
models, the no-black-hole model is radially biased in the central
region, but not enough to make a significant improvement to the fit to
the data. In other words, the black hole model is still a better fit even given
the maximum radial bias that the radial velocities can tolerate.
Presumably, including the proper motion will lead to an even poorer
fit for the no-black-hole model

To have the increase be caused by heavy remnants, there are two main
problems: having the required number of remnants in the first place, and having
all remnants well within the observed core with a very steep
density profile. The core radius of $\omega\,$Cen is 50\arcsec. The total
mass inside 50\arcsec\ is $8\times10^4\,M_{\odot}$. If the dispersion
increase seen inside 10\arcsec\ is due to remnants, then for a cluster
with a $r^{-2}$ profile all remnants would have to be inside
10\arcsec; having $4\times10^4\,M_{\odot}$ of material clustered inside
10\arcsec\ within a core of 50\arcsec\ would cause the cluster to
evaporate on very short timescales (Maoz 1998). Furthermore, the total
mass in heavy remnants (neutron stars and white dwarfs over $1\,M_{\odot}$)
would require a very top heavy initial mass function, inconsistent
with what is generally observed. The main problem with alternatives to
a black hole is that the velocity dispersion rises inside the core of
$\omega\,$Cen.

The black hole model fits the $\omega\,$Cen data the best and is consistent with an
extrapolation of the black hole $M_{{\rm bh}} - \sigma$ correlation. The same situation
is true in G1. However, for M15 and 47~Tuc, the black hole model is
preferred but not statistically significant. A main observational
point is that there is no black hole mass estimate for a GC
that is below the expected value from the $M_{{\rm bh}} - \sigma$
correlation.

\section{$N$-body Simulations of Massive Black Hole Formation}

The first $N$-body simulations of the formation of IMBHs in young,
dense clusters and their later interactions with
passing stars have been performed recently. In some cases, it is found that a massive
($>1000\, M_\odot$) object can form as a result of collisions between
young stars and that, if turning into an IMBH, this star will capture
passing stars through tidal energy dissipation. Gas accretion from circularized
stars onto the IMBH may be sufficient to create ultra-luminous X-ray sources 
(ULXs) in the cluster. 
These simulations therefore strengthen the connection between ULXs and IMBHs.

ULXs are point-like X-ray sources
with isotropic X-ray luminosities in excess of $L=10^{40}\,{\rm erg\,s}^{-1}$.
Since the Eddington luminosity of a star of mass $M$ is given by
$L_{\rm Edd}=1.3\times10^{38}\,{\rm erg\,s}^{-1}{(M/ M_\odot)}$,
where $M$ is the mass of the accreting object, most low-luminosity ULXs
are probably stellar-mass black holes. However, there is mounting
evidence that the brightest ULXs with luminosities exceeding
$10^{40}$\,erg\,s$^{-1}$ could be IMBHs.

The starburst galaxy M82 for example hosts a ULX with brightness in the range
$L=(0.5-1.6) \cdot 10^{41} {\rm erg\, s^{-1}}$
(\cite[Matsumoto \etal\ 2001]{metal01}, \cite[Kaaret \etal\ 2001]{ketal01}), corresponding to a black
hole with mass $350 -1200\,M_\odot$ if emitting photons at the
Eddington luminosity.  The case for an IMBH in M82 is supported by a
54\,mHz quasi-periodic oscillation found in the X-ray flux
(\cite[Strohmayer \& Mushotzky 2003]{sm03}) and also by the soft X-ray spectrum of this
source (\cite[Fiorito \& Titarchuk 2004]{ft04}). Additional observational
support for an IMBH comes from the
observation of a 62-day period in the X-ray luminosity
(\cite[Kaaret, Simet \& Lang 2006]{ksl06}, \cite[Patruno \etal\ 2006]{petal06}). 
The position of the ULX in M82 coincides 
with that of the young star cluster MGG-11. Recent
$N$-body simulations, summarized below, have showed how 
runaway merging of young stars could have led to the formation of
an IMBH in MGG-11 and how this IMBH later could have captured  passing
stars to became a ULX.

The evolution of MGG-11 was simulated through $N$-body simulations of
star clusters containing $N = 131,072$ (128K) stars using Aarseth's
collisional $N$-body code NBODY4 on the GRAPE computers in Bonn
and Tokyo (see Sec.~2). The initial set-up was given by King models
with various central concentrations in the range $3 \le W_0 \le 12$ 
and half-mass radius $r_h=1.3\,$pc. The initial mass function of cluster 
stars was given by a Salpeter power-law between 
$1.0\,M_\odot \le m \le 100\,M_\odot$. These clusters have a 
projected half-mass radius, mass-to-light ratio and total cluster 
mass after 12\,Myr (the age of MGG-11) that are consistent with the
properties of MGG-11 as observed by \cite{mgg03}.
In the simulations, stars were merged if their separation became smaller than the sum of
their radii. Orbital energy loss by tidal interactions 
between a star and the IMBH was implemented in the $N$-body simulations 
using the prescription of \cite{spzm93}. More details on these simulations
are presented in \cite{spz04} and \cite{b06}. 

For low-concentration models ($W_0 < 8$), the core radii hardly change with time
in the first few Myrs and expand at later times due to stellar
evolution mass loss from massive stars. Few stellar collisions are observed in
these models and no IMBH is formed. For clusters with higher concentration,
the central relaxation time is short enough that massive stars spiral into
the cluster core while still being in the hydrogen burning stage. 
Once in the cluster core, they can collide with each other due to the
high central density and the large stellar radii of massive
stars. Repeated collisions
lead to the formation of a VMS with $m>300\,M_\odot$.
Once such a VMS is produced, all further collisions are predominantly with this star
and its mass grows up to $m=500\,M_\odot - few \cdot 1000\,M_\odot$ (see also Sec.~3).
If this VMS collapses to an IMBH at the end of its lifetime,
the presence of a ULX in MGG-11 could be explained by this IMBH. Furthermore, simulations
of other young star clusters in M82 show that runaway merging of stars can  
happen only in MGG-11 and in none of the other clusters, explaining
why only MGG-11 hosts a ULX.

In another set of runs the further dynamical evolution of the IMBH in M82 was 
studied. These simulations show that a cusp develops in the stellar density distribution
around the IMBH. Inside this cusp, high-mass stars are enriched due to 
dynamical-friction-driven inspiral. Encounters of stars with the IMBH could
lead to tidal capture of the star if its pericenter distance is only
slightly larger than the tidal radius, $r_p/r_t \simlt 3$. If the orbit is
unperturbed by other stars, repeated pericenter
passages will further decrease its orbital semimajor axis until it
circularizes near the black hole. Angular momentum conservation requires this 
circularization to end at an orbital radius equal to twice the initial pericenter 
distance. Although
perturbations by other stars can either scatter the inspiraling star away from
the IMBH or onto an orbit with $r_p < r_t$ where it is destroyed, the simulations
show that on average $\approx 3$ successful inspirals leading to circularization 
happen within the lifetime of MGG-11.

Once circularized, stars will sooner or later fill their Roche lobe due to stellar
evolution and start to transfer mass onto the IMBH. The combined star-IMBH system
will then become visible as a ULX. In total, in 10 out of the 12 performed runs 
a ULX source was produced at least once between 3 and 12 Myrs. Furthermore, 4 
runs created an X-ray source brighter than $2\cdot10^{39}$ erg/sec 
within the age range of MGG-11. 
Hence, the performed $N$-body simulations provide a good explanation for the 
ULX source seen in MGG-11 (cf.\ Blecha \etal\ 2006).

A further test of this scenario might come from the stellar and orbital evolution of
the IMBH binaries. Since the runs show that mostly massive stars are captured and circularize
near the IMBH, stellar-mass black holes or NSs will be formed out of the
donor stars after they have undergone a supernova. The further evolution of
the IMBH binaries will then be driven by encounters with cusp stars, which 
harden the binaries. In the later stages, emission of gravitational waves will also
be important and will lead to the merger of the stars with the IMBH.
\cite{hp05} have shown that the event rate for this is likely to be high 
enough to be detectable by LISA, in particular if the IMBH mass is larger than
$\sim3\times10^3\,M_\odot$. Observations of gravitational waves from such
binaries would therefore give further support to the scenario discussed here.

\begin{acknowledgments}

G.~F.\ acknowledges support from NASA contract NAS8-39075 (CXC) and {\it Chandra\/} GO grant G06-7079A.
J.~M.~F.\ acknowledges support from NASA Grant NNG06GI62G and a {\it Chandra\/} Theory grant.
C.~O.~H.\ acknowledges support from the Lindheimer Fellowship at Northwestern University and thanks his many collaborators, especially J.\ Grindlay, P.\ Edmonds, and G.\ Rybicki.
S.~R.\ thanks the whole GBT cluster pulsar
search and timing team, especially I.~Stairs, J.~Hessels, P.~Freire,
and S.~Begin.
F.~A.~R.\ thanks Ingrid Stairs for help with the organization of IAU JD06, particularly
in Prague.

\end{acknowledgments}

\end{document}